\documentclass[a4paper,12pt]{article}
\usepackage[cp1251]{inputenc}
\usepackage[russian]{babel}
\usepackage{latexsym}
\usepackage{amsmath,amssymb}
\usepackage[active]{srcltx}
\begin{document}

\noindent
{\bf THE SIMPLE NONPOLAR CONTINUUM MEDIA.\\ PART III. THE
ASYMMETRIC ELASTICITY (N=2).}
 \bigskip
\\
{\bf V.O.Bytev}, Tyumen State University\\
 Tyumen, 625003, vbytev@utmn.ru
 \bigskip
 \\
Let $T$ be a stress tensor with components $\sigma _{ij} = \sigma
_{ji}$; $(i,j = 1,2)$. If a thin plate is loaded by forces at the
boundary, parallel to the to the plane of the plate and
distributed uniformly over the thickness the state of stress is
the specified by $\sigma _{11}$, $\sigma _{12}=\sigma _{21}$,
$\sigma _{22}$ only, and is called plane stress [1].

As it has been shown in [2] the most general form of $T$ can be
present as
\begin{equation} \label{EQB1}
 T = \theta \lambda _0 I + M \varepsilon ^0
\end{equation}
here
\begin{equation} \label{EQB2}
 M = \left( {\begin{array}{*{20}c}
   \mu  & {\mu _0 }  \\
   { - \mu _0 } & \mu   \\
\end{array}} \right);
\end{equation}
\begin{equation} \label{EQB3}
 \varepsilon ^0 = \left( {\begin{array}{*{20}c}
   u_x-\upsilon_y  & u_y+\upsilon_x  \\
   u_y+\upsilon_x & \upsilon_y-u_x   \\
\end{array}} \right);
\end{equation}
\begin{equation} \label{EQB4}
 \theta = div(\vec{u});
\end{equation}
$\vec{u}$ is the vector of the displacement; $\lambda _0$, $\mu$,
$\mu_0$ are the moduli of the elasticity, where
\begin{equation} \label{EQB5}
 \lambda _0 \equiv \lambda + \mu > 0;\qquad \mu > 0;
\end{equation}
$\mu_0$ is arbitrary parameter; $\sigma_{ij}$ are components of
the stress tensor $T$; $e^{ij} = e^{ji}$ are components of the
strain tensor;
\begin{equation} \label{EQB6}
 e^{ij} = \frac{1}{2}\left( \frac{\partial u^i}{\partial x^j} + \frac{\partial u^j}{\partial x^i}
 \right),\qquad (i,j = 1,2)
\end{equation}
and $u^1 \equiv u$, and $u^2 \equiv \upsilon$; $x^1 \equiv x$, and
$x^2 \equiv y$.

It can be shown in the usual way that matrix
$$
\frac{1}{{\mu_0}^2 + {\mu}^2} M
$$
is orthogonal. Therefore \eqref{EQB1} is a variant of wellknown
theorem of polar decomposition.

Using \eqref{EQB1} we get
$$
 \sigma_{11} = (\mu + \lambda_0) e^{11} + 2\mu_0 e^{12} + (\lambda_0 -
 \mu) e^{22}
$$
\begin{equation} \label{EQB7}
 \sigma_{12} = -\mu_0 e^{11} + 2\mu e^{12} + \mu_0
 e^{22}\qquad \qquad \quad
\end{equation}
$$
 \sigma_{22} = (\lambda_0 - \mu) e^{11} - 2\mu_0 e^{12} + (\lambda_0
 +
 \mu) e^{22}
$$

Let
\begin{equation} \label{EQB8}
 A = \left( {\begin{array}{*{20}c}
   {\mu + \lambda_0}  & {2\mu_0} & {\lambda_0 - \mu}  \\
   {-\mu_0} & {2\mu}  & {\mu_0} \\
   {\lambda_0 - \mu} & {-2\mu_0} & {\lambda_0 + \mu} \\
\end{array}} \right)
\end{equation}
is the matrix of system \eqref{EQB9}.
We obviously have
\begin{equation} \label{EQB9}
 \det A = 8 \lambda_0 ({\mu_0}^2 + {\mu}^2) > 0
\end{equation}
Hence, we have
\begin{equation} \label{EQB10}
 A^{-1} = \frac{1}{4 \lambda_0 ({\mu_0}^2 + {\mu}^2)} \left( {\begin{array}{*{20}c}
   {{\mu_0}^2 + {\mu}^2 + \lambda_0 \mu}  & {-2\mu_0 \lambda_0} & {{\mu_0}^2 + {\mu}^2 -\lambda_0 \mu}  \\
   {\mu_0 \lambda_0} & {2\mu \lambda_0}  & {-\mu_0 \lambda_0} \\
   {{\mu_0}^2 + {\mu}^2 - \lambda_0 \mu} & {2\mu_0 \lambda_0} & {{\mu_0}^2 + {\mu}^2 + \lambda_0 \mu} \\
\end{array}} \right)
\end{equation}
Therefore, from \eqref{EQB7} we get
$$
e^{11} = \frac{2}{\det A} \left[ ({\mu_0}^2 + {\mu}^2 + \lambda_0
\mu)\sigma_{11} - 2\mu_0 \lambda_0 \sigma_{12} + ({\mu_0}^2 +
{\mu}^2 - \lambda_0 \mu)\sigma_{22} \right]
$$
\begin{equation} \label{EQB11}
e^{12} = \frac{2}{\det A} \left[ \lambda_0 \mu_0 \sigma_{11} +
2\mu \lambda_0 \sigma_{12} -  \lambda_0 \mu_0 \sigma_{22} \right]
\qquad \qquad \qquad \qquad \qquad \quad
\end{equation}
$$
e^{22} = \frac{2}{\det A} \left[ ({\mu_0}^2 + {\mu}^2 - \lambda_0
\mu)\sigma_{11} + 2\mu_0 \lambda_0 \sigma_{12} + ({\mu_0}^2 +
{\mu}^2 + \lambda_0 \mu)\sigma_{22} \right]
$$

In the case of two-dimensional problems only three strain
components need be considered, namely
$$
e^{11} = \frac{\partial u}{\partial x},\quad e^{12} = \frac{1}{2}
\left( \frac{\partial u}{\partial y} + \frac{\partial
\upsilon}{\partial x} \right),\quad e^{22} = \frac{\partial
\upsilon}{\partial y}
$$
These three components are expressed by two functions $u$ and
$\upsilon$; hence they cannot be taken arbitrary. As it good
known, there exists a certain relations between the strain
components which can easily be written [1]
\begin{equation} \label{EQB12}
e^{11}_{yy} + e^{22}_{xx} = 2e^{12}_{xy}
\end{equation}

Using \eqref{EQB12}, we get
$$
({\mu_0}^2 + {\mu}^2 + \lambda_0 \mu)\triangle(\sigma_{11} +
\sigma_{22}) = {}
$$
$$
{} = 2\lambda_0\mu \left[ \frac{\partial}{\partial y} \left(
\frac{\partial \sigma_{22}}{\partial y} + \frac{\partial
\sigma_{12}}{\partial x} \right) + \frac{\partial}{\partial x}
\left( \frac{\partial \sigma_{11}}{\partial x} + \frac{\partial
\sigma_{12}}{\partial y} \right) \right] + {}
$$
$$
+ 2\lambda_0\mu_0 \left[ \frac{\partial}{\partial y} \left(
\frac{\partial \sigma_{11}}{\partial x} + \frac{\partial
\sigma_{12}}{\partial y} \right) - \frac{\partial}{\partial x}
\left( \frac{\partial \sigma_{22}}{\partial y} + \frac{\partial
\sigma_{12}}{\partial x} \right) \right]
$$
Taking into account the differential equations of equilibrium, we
obtain
$$
\frac{\partial \sigma_{11}}{\partial x} + \frac{\partial
\sigma_{12}}{\partial y} = 0; \quad \frac{\partial
\sigma_{12}}{\partial x} + \frac{\partial \sigma_{22}}{\partial y}
= 0
$$
\begin{equation} \label{EQB13}
\triangle(\sigma_{11}+\sigma_{22}) = 0
\end{equation}
$$
\sigma_{11} = 2\mu u_x + \mu_0(u_y+\upsilon_x) + (\lambda_0 -
\mu)\theta
$$
$$
\sigma_{12} = 2\mu \upsilon_y - \mu_0(u_y+\upsilon_x) + (\lambda_0
- \mu)\theta
$$
$$
\sigma_{22} = \mu(u_y+\upsilon_x) - \mu_0(u_x - \upsilon_y)\qquad
\quad
$$

Introducing function Airy $U$:
\begin{equation} \label{EQB14}
\sigma_{11} = U_{yy},\quad \sigma_{12} = -U_{xy},\quad \sigma_{22}
= U_{xx}
\end{equation}
we get
\begin{equation} \label{EQB15}
\triangle \triangle U = 0
\end{equation}

Besides, using \eqref{EQB13} and \eqref{EQB14}, we have
$$
(\mu + \lambda_0)u_x + 2\mu_0 (u_y + \upsilon_x)\frac{1}{2} +
(\lambda_0 - \mu) \upsilon_y = U_{yy}
$$
\begin{equation} \label{EQB16}
(\lambda_0 - \mu)u_x - 2\mu_0 (u_y + \upsilon_x)\frac{1}{2} +
(\lambda_0 + \mu) \upsilon_y = U_{xx}
\end{equation}
$$
-\mu_0 u_x + 2\mu (u_y + \upsilon_x)\frac{1}{2} + \mu_0 \upsilon_y
= -U_{xy}
$$

It easy shown that
$$
u_x = \frac{1}{4\lambda_0} (\sigma_{11} + \sigma_{22}) +
\frac{\mu}{4(\mu_0^2+\mu^2)} (\sigma_{11} - \sigma_{22}) -
\frac{2\mu_0}{4(\mu_0^2+\mu^2)} \sigma_{12};
$$
\begin{equation} \label{EQB17}
\upsilon_y = \frac{1}{4\lambda_0} (\sigma_{11} + \sigma_{22}) -
\frac{\mu}{4(\mu_0^2+\mu^2)} (\sigma_{11} - \sigma_{22}) +
\frac{2\mu_0}{4(\mu_0^2+\mu^2)} \sigma_{12};
\end{equation}
and
$$
\sigma_{11} + \sigma_{22} = \triangle U
$$

Let $$Q = \triangle U$$ then
\begin{equation} \label{EQB18}
\sigma_{11} = U_{yy} = Q - U_{xx}, \quad \sigma_{22} = U_{xx} = Q
- U_{yy}
\end{equation}

Using \eqref{EQB15} and \eqref{EQB18}, we obtain
\begin{equation} \label{EQB19}
\triangle Q = 0
\end{equation}
$$
\frac{\partial u}{\partial x} = - \frac{2\mu}{4(\mu_0^2+\mu^2)}
U_{xx} + \frac{2\mu_0}{4(\mu_0^2+\mu^2)} U_{xy} + \left[
\frac{1}{4\lambda_0} + \frac{\mu}{4(\mu_0^2+\mu^2)} \right] Q
$$
\begin{equation} \label{EQB20}
\frac{\partial \upsilon}{\partial y} = -
\frac{2\mu}{4(\mu_0^2+\mu^2)} U_{yy} -
\frac{2\mu_0}{4(\mu_0^2+\mu^2)} U_{xy} + \left[
\frac{1}{4\lambda_0} + \frac{\mu}{4(\mu_0^2+\mu^2)} \right] Q
\end{equation}

Taking into account \eqref{EQB19}, we get
$$
Q+iR = f(z), \quad z = x + iy;
$$
$$
\frac{\partial Q}{\partial x} = \frac{\partial R}{\partial y},
\quad \frac{\partial Q}{\partial y} = - \frac{\partial R}{\partial
x}.
$$
A function $f(z)$ is an analytic function of $z$. The integral of
this function with respect to $z$ is another analytic function.
Then writting $p$ and $q$ for real and imaginary parts of
$\varphi(z)$, we obtain
$$
\varphi(z) = p + iq = \frac{1}{4}\int f(z) dz
$$
so that
$$
\varphi^\prime (z) = \frac{1}{4} f(z)
$$
then
$$
\frac{\partial \varphi}{\partial z} = \frac{\partial p}{\partial
x} - i \frac{\partial p}{\partial y} = \frac{1}{4} (Q + iR)
$$

Using Cauchy--Rieman's condition, we have
$$
\frac{\partial p}{\partial x} = \frac{\partial q}{\partial y} =
\frac{1}{4} Q
$$
$$
\frac{\partial p}{\partial y} = -\frac{\partial q}{\partial x} =
\frac{1}{4} R
$$
therefore
\begin{equation} \label{EQB21}
Q = 4 \frac{\partial p}{\partial x} = 4 \frac{\partial q}{\partial
y}
\end{equation}

Thus any stress function is expressible in form
\begin{equation} \label{EQB22}
U = Re \left[ \overline{z} \varphi(z) + \chi(z) \right]
\end{equation}
or in another view [3]
$$
2U = \overline{z} \varphi (z) + \overline{\varphi (z)} + \chi (z)
+ \overline{\chi (z)}
$$
Using \eqref{EQB21}, we get
$$
2 \frac{\partial U}{\partial x} = \varphi(z) + \overline{z}
\varphi^\prime (z) + \overline{\varphi (z)} + z
\overline{\varphi^\prime (z)} + \chi^\prime (z) +
\overline{\chi^\prime (z)}
$$
$$
2 \frac{\partial U}{\partial y} = i \left[ -\varphi(z) +
\overline{z} \varphi^\prime (z) + \overline{\varphi (z)} - z
\overline{\varphi^\prime (z)} + \chi^\prime (z) -
\overline{\chi^\prime (z)} \right]
$$
therefor
$$
\frac{\partial U}{\partial x} + i \frac{\partial U}{\partial y} =
\varphi(z) + z \overline{\varphi^\prime (z)} + \overline{\psi (z)}
$$
where
$$
\psi(z) = \frac{d\chi}{dz}
$$

Substituting \eqref{EQB21} to \eqref{EQB20}, we get
\begin{equation} \label{EQB23}
\frac{\partial u}{\partial x} = \left( \frac{1}{\lambda_0} +
\frac{\mu}{\mu_0^2 + \mu^2} \right) \frac{\partial p}{\partial x}
- \frac{\mu}{2(\mu_0^2 + \mu^2)} U_{xx} + \frac{\mu_0}{2(\mu_0^2 +
\mu^2)} U_{xy}
\end{equation}
\begin{equation} \label{EQB24}
\frac{\partial \upsilon}{\partial y} = \left( \frac{1}{\lambda_0}
+ \frac{\mu}{\mu_0^2 + \mu^2} \right) \frac{\partial q}{\partial
x} - \frac{\mu}{2(\mu_0^2 + \mu^2)} U_{yy} -\frac{\mu_0}{2(\mu_0^2
+ \mu^2)} U_{xy}
\end{equation}
Integrating \eqref{EQB23} with respect to $x$, and \eqref{EQB24}
with respect to $y$, we obtain
\begin{equation} \label{EQB25}
u = \frac{\mu_0^2 + \mu^2 + \lambda_0 \mu}{\lambda_0(\mu_0^2 +
\mu^2)} p - \frac{\mu}{2(\mu_0^2 + \mu^2)} U_{x} +
\frac{\mu_0}{2(\mu_0^2 + \mu^2)} U_{y} + f_1(y)
\end{equation}
\begin{equation} \label{EQB26}
\upsilon = \frac{\mu_0^2 + \mu^2 + \lambda_0
\mu}{\lambda_0(\mu_0^2 + \mu^2)} q - \frac{\mu}{2(\mu_0^2 +
\mu^2)} U_{y} - \frac{\mu_0}{2(\mu_0^2 + \mu^2)} U_{x} + f_2(x)
\end{equation}
Using \eqref{EQB16}, we have
\begin{equation} \label{EQB27}
u_y + \upsilon_x = \frac{\mu_0}{2(\mu_0^2 + \mu^2)} (U_{yy} -
U_{xx}) - \frac{\mu}{\mu_0^2 + \mu^2} U_{xy}
\end{equation}
but from \eqref{EQB25} and \eqref{EQB26}, we get
\begin{equation} \label{EQB28}
u_y + \upsilon_x = \frac{\mu_0^2 + \mu^2 + \lambda_0
\mu}{\lambda_0(\mu_0^2 + \mu^2)} (p_y + q_x) +
\frac{\mu_0}{2(\mu_0^2 + \mu^2)} (U_{yy} - U_{xx}) -
\frac{\mu}{\mu_0^2 + \mu^2} U_{xy} + \frac{df_1(y)}{dy} +
\frac{df_2(x)}{dx}
\end{equation}
Comparison \eqref{EQB27} and \eqref{EQB28} is given by equation
$$
\frac{df_1(y)}{dy} + \frac{df_2(x)}{dx} = 0.
$$
This implies that
$$
f_1(y) = s_1(-\varepsilon y + \alpha_1)
$$
\begin{equation} \label{EQB29}
f_2(x) = s_1(\varepsilon x + \alpha_2)
\end{equation}

Substituting \eqref{EQB29} in \eqref{EQB25} and \eqref{EQB26}, we
have without loss of generality that
\begin{equation} \label{EQB30}
u = \frac{\mu_0^2 + \mu^2 + \lambda_0 \mu}{\lambda_0(\mu_0^2 +
\mu^2)} p - \frac{\mu}{2(\mu_0^2 + \mu^2)} U_{x} +
\frac{\mu_0}{2(\mu_0^2 + \mu^2)} U_{y}
\end{equation}
\begin{equation} \label{EQB31}
\upsilon = \frac{\mu_0^2 + \mu^2 + \lambda_0
\mu}{\lambda_0(\mu_0^2 + \mu^2)} q - \frac{\mu_0}{2(\mu_0^2 +
\mu^2)} U_{y} - \frac{\mu}{2(\mu_0^2 + \mu^2)} U_{x}
\end{equation}
here
$$
p = p(x,y);\quad q = q(z,y)
$$

Using \eqref{EQB30} and \eqref{EQB31}, we get
\begin{equation} \label{EQB32}
u + i \upsilon = \frac{\mu_0^2 + \mu^2 + \lambda_0
\mu}{\lambda_0(\mu_0^2 + \mu^2)} (p + iq) - \frac{\mu +
i\mu_0}{2(\mu_0^2 + \mu^2)} (U_{x} + iU_{y})
\end{equation}

Let
\begin{equation} \label{EQB33}
{\ae} = \mu + i\mu_0, \quad |{\ae}|^2 = \mu_0^2 + \mu^2
\end{equation}
taking into account that
$$
p + iq = \varphi(z)
$$
and
$$
\triangle \triangle U = 0,
$$
we obtain
\begin{equation} \label{EQB34}
|{\ae}|^2 (u + i\upsilon) = \frac{2}{\lambda_0} \left( |{\ae}|^2 +
\lambda_0 Re({\ae}) \right) \varphi (z) - {\ae} \left[ \varphi (z)
+ z \overline{\varphi^\prime (z)} + \overline{\psi (z)} \right]
\end{equation}
Substituting $\mu_0 = 0$ in \eqref{EQB34}, we get classical
Loeve's formula (see, for example, [1]).
$$
2\mu (u + i\upsilon) = {\ae}_0 \varphi(z) - z
\overline{\varphi^\prime (z)} - \overline{\psi (z)};
$$
$$
{\ae}_0 = \frac{\lambda + 3 \mu}{\lambda + \mu}.
$$

As for components of the stress tensor, we have classical formula
[1]
$$
\sigma_{11} + \sigma_{22} = 2 [ \varphi^\prime (z) +
\overline{\varphi^\prime (z)} ] = 4 Re [\varphi^\prime (z)],
$$
$$
\sigma_{22} - \sigma_{11} + 2 i \sigma_{12} = 2 [\overline{z}
\varphi^{\prime \prime} (z) + \psi^{\prime \prime} (z) ].
$$

\textit{Example.} The effect of circular holes on stress
distributions and displacements in plates (see, for example, [1]).

Let us assume that two complex lanes $Z$ and $G$ are given and
conformal mapping $z = w(\zeta)$ of area $S \subset Z$ to area
$\Sigma \subset G$ is defined. Now let us introduce polar
coordinates $(r,\theta)$ at $G$ plane, so that $w(\zeta) = \zeta =
re^{i\theta}$. Then any vector $(w_x,w_y)$ is transformed in
compliance to the following formula: $w_r + i w_\theta =
e^{-i\theta}(w_x + w_y)$. With help of this we can get polar
representation of translation vector $w = (u_r,u_\theta)$ from
\eqref{EQB34}

\begin{equation} \label{EQB35}
|{\ae_0}|^2 (u_r + iu_\theta) = e^{-i\theta} \left(
(2\lambda_0^{-1} {\ae}_0^2 + \overline{{\ae}})\varphi(\zeta) -
{\ae} \zeta \overline{\varphi^{\prime}(\zeta)} -
{\ae}\overline{\psi(\zeta)} \right)
\end{equation}

Let a plate submitted to a uniform tension of magnitude $p$ in the
$x$--direction. If a small circular hole is made in the middle of
the plate, the stress distribution in the neighborhood of the hole
will be changed, but we can conclude from Saint--Venant's
principle that the change is negligible at distances with are
large compared with $R$, the radius of the hole.

Let us assume that the edges of the plate are free from external
stress and at infinity $\sigma_{11}^{\infty}=p$,
$\sigma_{22}^{\infty}=0$, $\sigma_{12}^{\infty}=0$, that means
that stress is present along $0x$ axis, tensile stress at infinity
being a constant value $p$. In this case functions $\varphi(z)$
and $\psi(z)$ and tensor components shall be defined with the help
of the following equations [4].

As usually [1], we obtain
$$
\sigma_{rr} = \frac{p}{2} \left( 1 - \frac{R^2}{r^2} \right) +
\frac{p}{2} \left( 1 - \frac{4 R^2}{r^2} + \frac{3 R^4}{r^4}
\right) \cos (2\theta)
$$
\begin{equation} \label{EQB36}
\sigma_{\theta \theta} = \frac{p}{2} \left( 1 + \frac{R^2}{r^2}
\right) - \frac{p}{2} \left( 1 + \frac{3 R^4}{r^4} \right) \cos
(2\theta)
\end{equation}
$$
\sigma_{r \theta} = -\frac{p}{2} \left( 1 + \frac{2 R^2}{r^2} -
\frac{3 R^4}{r^4} \right) \sin (2\theta)
$$

The functions $\varphi(z)$ and $\psi(z)$ are
$$
\varphi(z) = \frac{p}{4} \left( z + \frac{2 R^2}{z} \right);
$$
$$
\psi(z) = - \frac{p}{2} \left( z + \frac{R^2}{z} - \frac{R^4}{z^3}
\right).
$$

Relevant \eqref{EQB35} translation vector components are as
follows:

$$
u_r = \frac{1}{4}pR \left( \lambda_0^{-1} \frac{r}{R} + \mu
{\ae}_0^{-2} \frac{R}{r} \right) + \frac{1}{4} \mu_0 {\ae}_0^{-2}
p R \left( \frac{r}{R} - 2 \frac{R}{r} + \frac{R^3}{r^3} \right)
\sin(2\theta) + {}
$$
$$
{} + \frac{1}{4}pR \left[ 2\lambda_0^{-1} \frac{R}{r} + \mu
{\ae}_0^{-2} \left( \frac{r}{R} + 2 \frac{R}{r} - \frac{R^3}{r^3}
\right) \right] \cos(2\theta),
$$
$$
u_{\theta} = \frac{1}{4} \mu_0 {\ae}_0^{-2} pR \left(\frac{R}{r} -
\frac{t}{R} \right) + \frac{1}{4} \mu_0 {\ae}_0^{-2} p R \left(
\frac{r}{R} - \frac{R^3}{r^3} \right) \cos(2\theta) + \qquad
\qquad {}
$$
$$
{} + \frac{1}{4}pR \left[ 2\lambda_0^{-1} \frac{R}{r} + \mu
{\ae}_0^{-2} \left( \frac{r}{R} + \frac{R^3}{r^3} \right) \right]
\sin(2\theta).
$$

Assuming $\mu_0 = 0$, we get traditional formulas of this problem
[1]:

\begin{equation} \label{EQB37}
u^{cl}_r = \frac{1}{4}pR \left( \lambda_0^{-1} \frac{r}{R} +
\mu^{-1}\frac{R}{r} \right) + \frac{1}{4}pR \left[
2\lambda_0^{-1}\frac{R}{r} + \mu_{-1} \left( \frac{r}{R} +
2\frac{R}{r} - \frac{R^3}{r^3} \right) \right] \cos(2\theta),
\end{equation}
\begin{equation} \label{EQB38}
u^{cl}_{\theta} = - \frac{1}{4}pR \left[
2\lambda_0^{-1}\frac{R}{r} + \mu_{-1} \left( \frac{r}{R} +
 \frac{R^3}{r^3} \right) \right] \sin (2\theta).
\end{equation}

Now let us calculate boundary translation values $r=R$ with help
of \eqref{EQB37} and \eqref{EQB38}:
$$
u_r|_{r=R} = \frac{pR (\mu_0^2 + 2 \mu^2 + \lambda \mu)}{4(\lambda
+ \mu)(\mu_0^2 + \mu^2)}(1 + 2\cos(2\theta)),
$$
$$
u_{\theta}|_{r=R} = - \frac{pR (\mu_0^2 + 2 \mu^2 + \lambda
\mu)}{2(\lambda + \mu)(\mu_0^2 + \mu^2)}\sin(2\theta).
$$

Passage to the limit where $\mu \rightarrow 0$ will give final
values boundary translations.

$$
u_r|_{r=R} \rightarrow \frac{pR}{4\lambda}(1 + 2\cos(2\theta)),
$$
$$
u_{\theta}|_{r=R} \rightarrow - \frac{pR}{2\lambda}\sin(2\theta).
$$

Traditional formulas will give the following values:
$$
u_r^{cl}|_{r=R} = \frac{pR (\lambda + 2\mu)}{4 \mu (\lambda +
\mu)}(1 + 2\cos(2\theta)),
$$
$$
u_{\theta}^{cl}|_{r=R} = - \frac{pR (\lambda + 2\mu)}{2 \mu
(\lambda + \mu)}\sin(2\theta),
$$
having no physical sense where $\mu \rightarrow 0$.

\bigskip

\textit{Remark.} The experiments has proved the validity of this
theory.


\newpage
\textit{Referenses.}

\begin{enumerate}
\item Timoshenko S., and Goodier J.N., 1951, "Theory of Elasticity". NY,
Toronto, London, McGraw--Hill Book Company, Inc.

\item Bytev V.O., 2009, "The Simple Nonpolar Continuum Media". Part I. The equivalence transformation. (archived article,  www.arxiv.org)

\item Muschelisvili N.I., 1932, Math.Ann., vol.~107,
pp.~282--312.

\item Muschelisvili N.I., "Some Mathematical Problems in the
Theory of Elasticity", Nauka, 1966. 642~p. (In Russian).
\end{enumerate}
  \bigskip

V.O. Bytev, Tyumen, e-mail: vbytev@utmn.ru

\end{document}